# RISTRETTO: a pathfinder instrument for exoplanet atmosphere characterization

Bruno Chazelas[a], Christophe Lovis[a], Nicolas Blind[a], Jonas Kühn[a], Ludovic Genolet[a], Ian Hughes[a], Martin Turbet[a], Janis Hagelberg[a], Nathanaël Restori[a], Markus Kasper[b], Nelly Natalia Cerpa Urra[b,c]

[a]Observatoire de Genève, University of Geneva, 51 chemin des maillettes 1290 Versoix, Switzerland
[b]European Southern Observatory, Karl-Schwarzschild-Str. 2, 85748 Garching, Germany
[c]Institut d'Optique Graduate School, 2 avenue Augustin Fresnel 91127 Palaiseau, France

## ABSTRACT

We introduce the RISTRETTO instrument for ESO VLT, an evolution from the original idea of connecting the SPHERE high-contrast facility to the ESPRESSO spectrograph[1]. RISTRETTO is an independent, AO-fed spectrograph proposed as a visitor instrument, with the goal of detecting and characterizing nearby exoplanets in reflected light for the first time. RISTRETTO aims at characterizing the atmospheres of Proxima b and several other exoplanets using the technique of high-contrast, high-resolution spectroscopy. The instrument is composed of two parts: a front-end to be installed on VLT UT4 providing a two-stage adaptive optics system, using the AOF facility as a first stage, with coronagraphic capability and a 7-fiber IFU, and a diffraction-limited R=135,000 spectrograph in the 620-840 nm range. We present the requirements and the preliminary design of the instrument.

## 1. INTRODUCTION

Direct detection of the light from extrasolar planets is a challenging objective. For young exoplanets on wide orbits this is achieved by high-contrast imaging and coronagraphy[2]. For close-in exoplanets it is possible to take advantage of planetary transits[3,4] and obtain a direct measurement of the IR flux of giant planets using secondary eclipses and phase curves. However, many known exoplanets including those orbiting the brightest and nearest stars remain out of reach of these techniques.

The RISTRETTO instrument will attempt reflected-light observations of spatially-resolved exoplanets for the first time. The proposed[1] technique for RISTRETTO is high-contrast, high-resolution (HCHR) spectroscopy[5–8]. To separate the planet from the star, adaptive optics on an 8m class telescope is used. In order to reach the required contrast, a coronagraph and an extreme AO stage are needed. Using a high-resolution spectrograph, the radial velocity shift between star and planet allows us to separate the stellar and planetary spectral lines, providing an additional factor ~1000 in achievable contrast. Other experiments are being developed based on a similar approach[9–11]. However they are targeting the thermal emission of young, massive planets at larger separations in the IR. Driven by the Proxima b science case, the RISTRETTO instrument will be targeting reflected light in the visible wavelength range. The short wavelengths allow us to spatially resolve Proxima b and other known very nearby exoplanets, placing them at about 2 λ/d on a 8m-class telescope.

Figure 1 shows the sample of known exoplanets as a function of their angular separation from their host star and their estimated contrast. RISTRETTO will target the most favorable objects, which have contrasts of ~$10^{-7}$ or above. Detecting these objects with RISTRETTO will constrain their albedo, orbital inclination and true mass. In addition, this technique has the potential to perform an in-depth spectral characterization of planetary atmospheres. Ultimately, when used at the ELT, it could detect O2, H2O, CH4 or CO2 in the atmosphere of the habitable-zone planet Proxima b. Figure 2 shows the estimated observing time needed to detect a number of known exoplanets with RISTRETTO, ranging from Jupiter-like gas giants to super-Earths and even the temperate rocky planet Proxima b. We expect that additional suitable planets will be discovered in the future by radial velocity surveys such as ESPRESSO, CARMENES, NIRPS, and SPiROU.

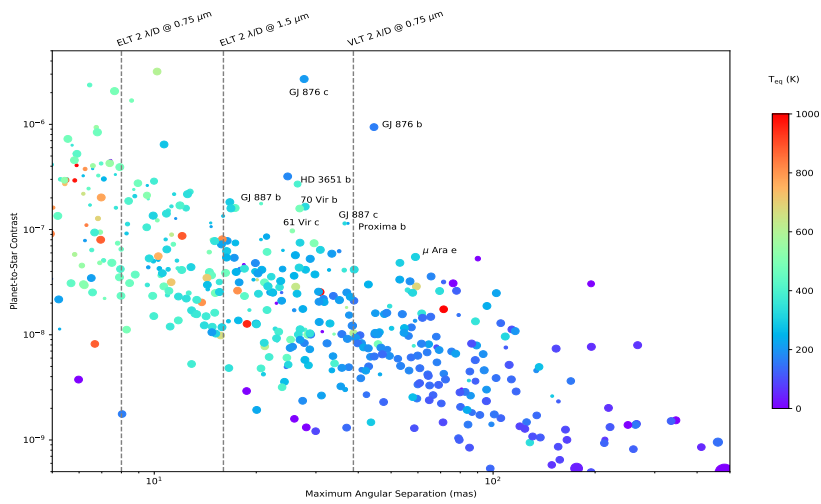

Figure 1: Planets represented in the star separation (at the maximum elongation) / contrast diagram. The labeled planets are within the reach of RISTRETTO.

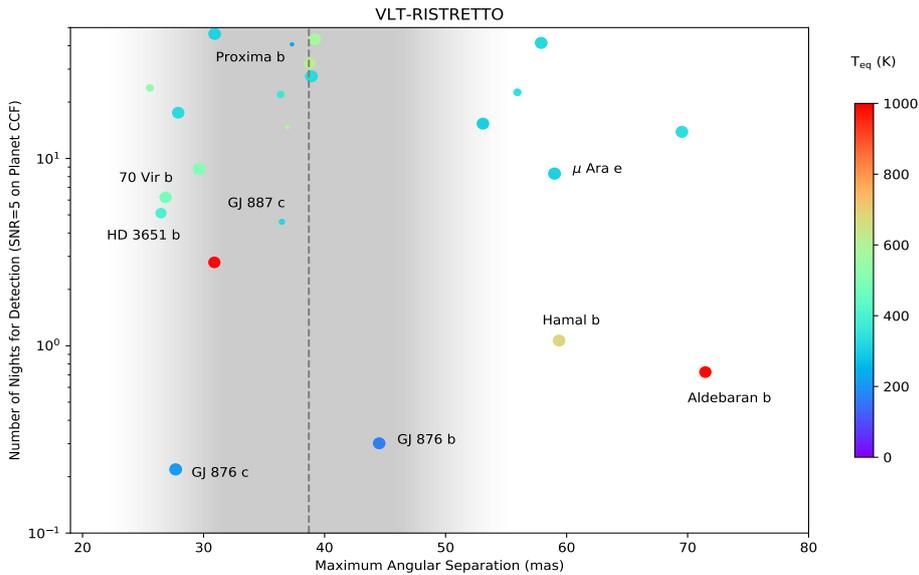

Figure 2: Integration time to detect the reflected light of different planets. A handful are well within RISTRETTO reach. Proxima b is challenging but could be be accessible to RISTRETTO depending on its bond albedo at maximum elongation[1] which depends on the nature of the atmosphere and surface of the planet[29]. It is to be noted that planets further away than 2λ/d can be observed. What is plotted here is the angular separation at the maximum elongation. At the appropriate orbital phase ,when a planet is observed observed closer to superior conjunction the reflected flux increases.

## 2. GLOBAL ARCHITECTURE OF THE INSTRUMENT

The first idea when designing this experiment was to link directly SPHERE to the ESPRESSO spectrograph. This however is difficult for different reasons. First SPHERE would need an important upgrade in the visible and the SPHERE+ upgrade is only for the infrared. Also the connection to ESPRESSO is not obvious, as one would probably need to add motors into the instrument (which would not be acceptable for an instrument with such high stability requirements). Thus we decided to start from a blank page. This allows to use a diffraction limited setup from the top to the bottom of the instrument, which is precious to achieve the required level of contrast.

The instrument will have 2 major components:

- A front-end feeding a 7-spaxels IFU, which includes an extreme AO (using UT4 AO Facility as a first stage) with a coronograph.
- A stable diffraction limited spectrograph having 7 input fibers.

The geometry of the spaxels on the sky is shown on figure 3, This geometry and the use of single mode fibers means that the full ring at 2 λ/d is not covered in a single exposure. Thus one needs 2 exposure with a 30° offset to cover the full ring around the star.

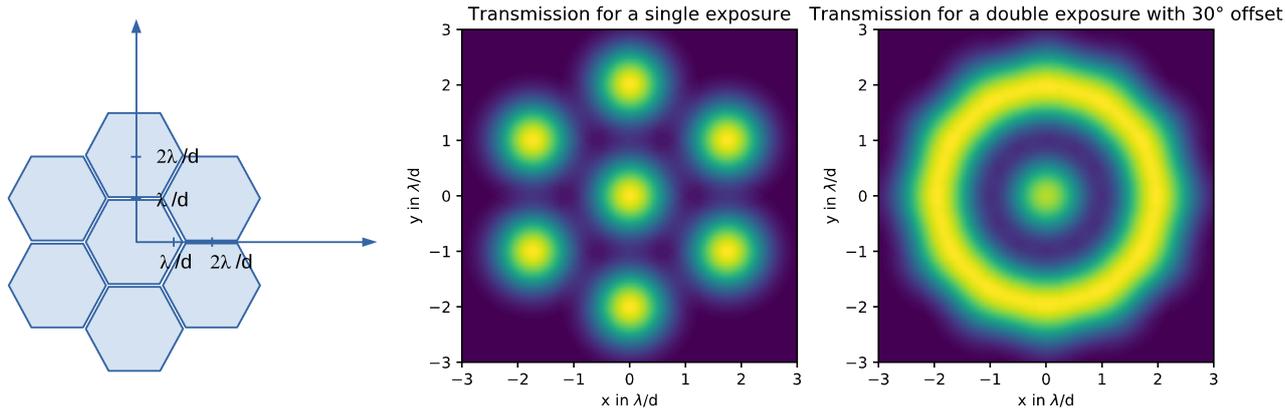

Figure 3: (left) disposition of the spaxel on the sky, (center) transmission of the IFU with one exposure, (right) transmission after 2 exposures with a 30° offset

## 3. DEVELOPMENT STRATEGY

Given our previous experience we decided that we would first develop the spectrograph. This comes with the opportunity to test it directly on the EULER telescope that is being equipped with a small adaptive optic system (KalAO[12]). This will let more time to develop the most challenging part of the experiment: the front-end. During this test phase where RISTRETTO will be mounted on the 1-m Euler telescope, we plan to carry out scientific observation campaigns on several solar system objects. Observations will include (but not be limited to) localized wind measurements[13–15], mapping of column-integrated and possibly altitude-dependent gas mixing ratios (e.g., of $O_2$, $H_2O$ and possibly $CH_4$) of the atmospheres of Venus, Mars, Titan, Jupiter and Saturn and possibly other solar system objects.

The final position of the instrument would be on VLT UT4 with AOF as a first stage. When the design will be advanced enough this experiment will be proposed as visitor instrument. The time frame for its installation would be in the interval between HAWK-I and MAVIS.

## 4. OPTICAL FIBER

In order to link the AO system and the spectrograph the natural solution is to use single mode fibers. It happens there are available standard single mode fiber for our exact wavelength range. S-630-HP from (NuFern or Thorlabs).

The other possibility was to look at PCF fibers. The have the advantage of offering a larger bandwidth but the currently available fibers have a blue cut-off that is really near to the blue-end of our scientific band-pass. Moreover their transmission is also not as good as the standard circular single mode fibers which will be problematic for RISTRETTO.

The option of using multi-core fibers is also interesting, however because this is not a mature technology in the visible we are first investigating the standard fiber case.

# 5. SPECTROGRAPH DESIGN

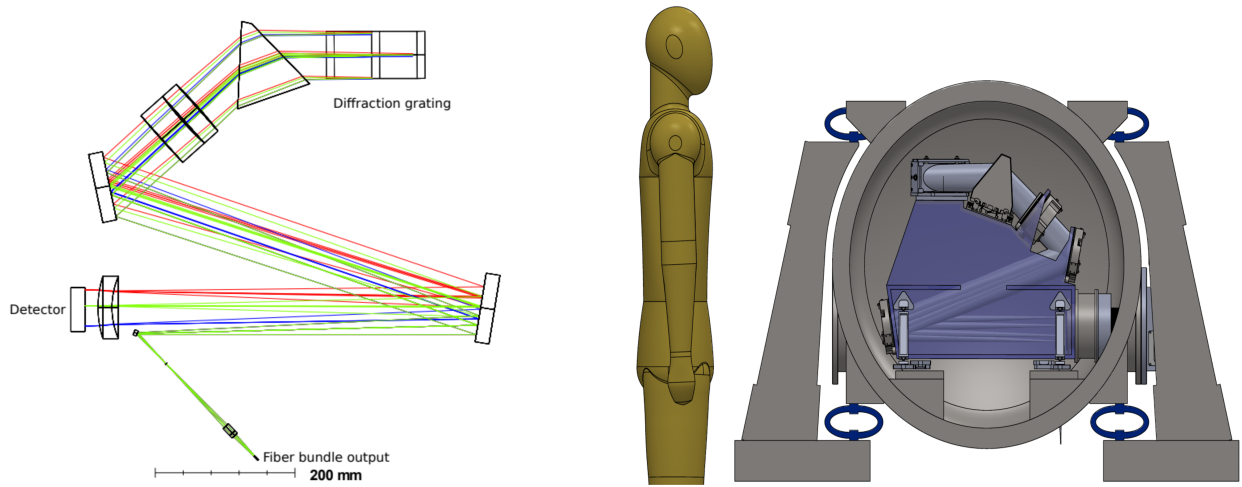

Figure 4: Design of the RISTRETTO spectrograph

There are already a number of design[16–21] of diffraction limited spectrographs, as this a tempting solution for radial-velocity machines. The main limitation, at least in the visible has been so far the need for an efficient AO system. This is one of the point we plan to address with this project.

Given the requirement in resolution, band-pass and the number of spaxel needed, we chose to keep a traditional echelle spectrograph design.

The spectrograph requirements are the following:

- 7 is the minimum number of fibres.
- Resolution has to be >= 130'000 goal 150'000
- Line positions should be stable on the detector at 1/100th of a pixel over 24 h
- Line Spread Function should be stable over 24h
- At a given wavelength one should be able to subtract a spectrum from an other spaxel without having LSF residual bigger than the noise (or we are able to find a way to calibrate the stellar spectrum for each spaxels).
- Sampling ~ 2.5 (> 2) pixels in the dispersion direction
- Minimum sampling in the cross dispersion direction 2.0 pixels (avoid under sampling)
- Band-pass 620 nm-840 nm (to cover the band for oxygen, water and Hα)
- The spectrograph throughput should be >= 40%

The minimum requirement for the detector is 4Kx4K. The detector selected will be a CCD-231-84 from E2V-teledyne.

An off the shelf grating from NEWPORT (53019ZD06-412E) is the heart of the spectrograph. It is a 63° blaze grating with a groove period of 23.2 mm$^{-1}$ and has a very good efficiency. The design is a double pass white pupil echelle spectrograph.

In such a spectrograph the distance between the input fibers is not a free parameter. In our case, separating them enough on the detector, in the cross dispersion direction would require the fiber to be nearer to one another than the standard cladding diameter of 125 microns allows. However only the projected vertical distance between the fibers is responsible for the spaxel orders separation: thus the solution is use the second dimension by strongly tilting the fiber "slit". One other possibility would be to use multi-core fibers, but as already stated our baseline is to use standard graded index fibers. The price to pay with such a tilted slit is that, in the red part of the spectrum there is some vignetting affecting differently the different fiber spectra. There are no holes though, thanks to the overlapping wavelength from one order to the other. An advantage of the long-tilted slit is to limit the interference patterns between the spaxels.

The resolution of the instrument is driven on one side by the dispersion of the grating and by the diffraction limit of the grating (contrarily to a standard fiber-fed spectrograph). The grating is 140x60 mm. We followed and confirmed an analysis[22] showing that you need to increase the grating size by a certain factor to get the diffraction limit, with a Gaussian beam, other wise you cut the wings of the Gaussian and increase the PSF size. We chose a factor 1,7.

To make a compact design one uses 2 fold mirrors so that the whole instrument can fit in an 800 mm diameter vacuum enclosure.

The rest of the design is very similar to HARPS or similar instruments. The grating is set to have the lines vertical.

One undesirable feature of such a spectrograph is that the PSF size is varying with wavelength (see figure 5 left panel). Given the wavelength range this means either we accept loosing resolution on the red side of the spectrum or we accept to be under-sampled on the blue side of the spectrum. The solution to this issue is using tilt to make a controlled chromatic defocus of the orders in order to get a more uniform PSF across the orders (see center panel on figure 5). The PSF will however not be as similar as we could require to directly compare the spectrum from the star from the spectrum of the planet. Thus it will probably be necessary to have a calibration of the stellar spectrum on each channel.

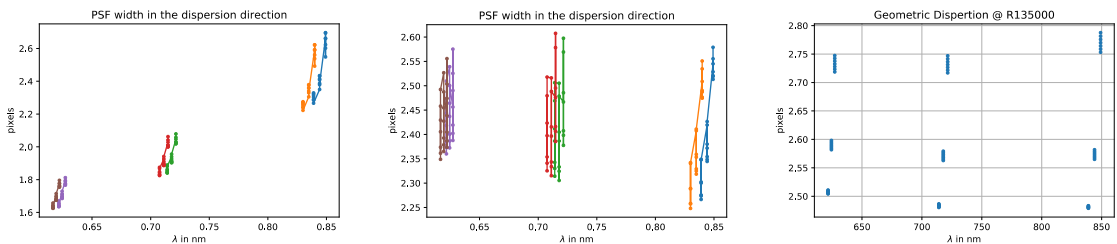

*Figure 5: Resolution of the spectrograph. On the left, the PSF FWHM with an in focus detector. In the center the PSF FWHM with a defocused and tilted detector. On the right the geometrical dispersion.*

The detector will be cooled using a Pulse tube cryo-cooler, using a differential vacuum design, similar to the one of HARPS or ESPRESSO.

## 6. IFU PRELIMINARY CONCEPTS

The IFU will have several functions:
- collect the light for the spectrograph
- provide the field derotation, and the offset needed to observe the full ring around the star.
- Provide some way of guiding.

Once again a solution with multi-core fiber is attractive, they have the natural geometry and it is possible to print directly micro lens on top of them. However given our wavelength-range. We are studying a solution using standard micro-lenses glued to standard optical fibers.

It is possible to have 2 different injection schemes in the fibers. Injection in the image plane or in the pupil plane. It the first plane, apart from pure transmission issue, the fiber/micro-lens alignment translates directly in offset of the on sky sensitivity window. We prefer the pupil plane alignment scheme. It is simpler and alignment requirement drive only the contrast and transmission. Zero order calculation show that alignment has to be done to the sub-micron regime. This is a challenge.

We think of using the waffle mode of the deformable mirror to provide for some guiding references.

## 7. CORONOGRAPH

There are 2 ways of looking at the coronograph:

- One is trying to have a low amount of light in the 2 λ/d ring,
- Or trying to do nulling[23].

The 2 approaches are under evaluation. We are looking at different approach, such as pupil plane devices. The nulling approach is very compelling as it might give additional degrees of freedom in the optimization. We are looking at solution that are polarization independent.

Note that the contrast will be ultimately limited by the performance of the AO system. Thus is not necessary to get a contrast better than $10^{-4}$ with the coronograph. Instead it is worth to try to improve the transmission to a maximum. One of the approach is to look into the possibility of the PIAA[24], as this is a compelling solution to increase the coupling into the fiber.

Another important open question: should we have a system that removes the light from the central spaxel? There is a high risk that the intense light in this spaxels would spill into the other spaxels via spurious reflections, diffraction by the micro-lens borders etc. On the other end if we manage to get this under control the system in the end would be much simpler as one could avoid an additional image plane.

## 8. FRONT END PRELIMINARY CONCEPTS

It seems that to achieve our goals on possible route is to have a system similar to what has been proposed for SPHERE+, or working on instrument such as SCEXAO[25]: having a 2 stage AO. In our case the system under study would be the following:

- The first stage would be the AOF[26] facility: a deformable secondary, with a Shack Hartman wavefront sensor. Sensing would be performed in the J band, running at 1 Khz.
- A second stage would have a pyramid wavefront sensor in the I band (minus the science bandwidth), a 24x24 Boston deformable mirror and work at high speed (at least 3.5 kHz)

The goal of the second stage will be to reduce the residuals of the low order aberrations that are the one that have the larger impact on the fiber coupling efficiency as well as on the contrast at 2λ/d in the fiber.

## 9. CONCLUSION

RISTRETTO is a promising experiment, with a clear science-case and can be seen as a pathfinder experiment in terms of science and technology to future ELT instruments such as HIRES[27] (and in particular its SCAO module), and PCS[28].

## 10. ACKNOWLEDGMENTS

We would like to thanks in particular Bernard Delabre that helped us with the spectrograph design concept.

This work has been carried out in part within the framework of the NCCR PlanetS supported by the Swiss National Science Foundation. The RISTRETTO project is partially funded through SNSF's FLARE programme for large research infrastructures.

## 11. REFERENCES

[1]   Lovis, C., Snellen, I., Mouillet, D., Pepe, F., Wildi, F., Astudillo-Defru, N., Beuzit, J.-L., Bonfils, X., Cheetham, A., Conod, U., Delfosse, X., Ehrenreich, D., Figueira, P., Forveille, T., Martins, J. H. C., Quanz, S. P., Santos, N. C., Schmid, H.-M., Ségransan, D., et al., "Atmospheric characterization of Proxima b by coupling the SPHERE high-contrast imager to the ESPRESSO spectrograph," Astronomy & Astrophysics **599**, A16 (2017).

[2] Pueyo, L., "Direct Imaging as a Detection Technique for Exoplanets," [Handbook of Exoplanets], H. J. Deeg and J. A. Belmonte, Eds., Springer International Publishing, Cham, 705–765 (2018).
[3] Alonso, R., "Characterization of Exoplanets: Secondary Eclipses," [Handbook of Exoplanets], H. J. Deeg and J. A. Belmonte, Eds., Springer International Publishing, Cham, 1–26 (2018).
[4] Parmentier, V. and Crossfield, I. J. M., "Exoplanet Phase Curves: Observations and Theory," [Handbook of Exoplanets], H. J. Deeg and J. A. Belmonte, Eds., Springer International Publishing, Cham, 1–22 (2017).
[5] Sparks, W. B. and Ford, H. C., "Imaging Spectroscopy for Extrasolar Planet Detection," The Astrophysical Journal $578$, 543–564 (2002).
[6] Riaud, P. and Schneider, J., "Improving Earth-like planets' detection with an ELT: the differential radial velocity experiment," Astronomy and Astrophysics $469$, 355–361 (2007).
[7] Snellen, I. A. G., Brandl, B. R., de Kok, R. J., Brogi, M., Birkby, J. and Schwarz, H., "Fast spin of the young extrasolar planet β Pictoris b," Nature $509$, 63–65 (2014).
[8] Snellen, I., de Kok, R., Birkby, J. L., Brandl, B., Brogi, M., Keller, C., Kenworthy, M., Schwarz, H. and Stuik, R., "Combining high-dispersion spectroscopy with high contrast imaging: Probing rocky planets around our nearest neighbors," Astronomy and Astrophysics $576$, A59 (2015).
[9] Vigan, A., "HiRISE: Bringing high-spectral," 39 (2019).
[10] Vigan, A., Otten, G. P. P. L., Muslimov, E., Dohlen, K., Philipps, M. W., Seemann, U., Beuzit, J.-L., Dorn, R., Kasper, M., Mouillet, D., Baraffe, I. and Reiners, A., "Bringing high-spectral resolution to VLT/SPHERE with a fiber coupling to VLT/CRIRES+," 1070236 (2018).
[11] "REACH — REACH documentation.", <http://secondearths.sakura.ne.jp/reach/index.html> (20 February 2020 ).
[12] Janis Hagelberg et al., "KalAO the swift adaptive optics imager on 1.2m Euler Swiss telescope in La Silla, Chile.," this conference, SPIE.
[13] Machado, P., Widemann, T., Luz, D. and Peralta, J., "Wind circulation regimes at Venus' cloud tops: Ground-based Doppler velocimetry using CFHT/ESPaDOnS and comparison with simultaneous cloud tracking measurements using VEx/VIRTIS in February 2011," Icarus $243$, 249–263 (2014).
[14] Machado, P., Silva, M., Peralta, J., Luz, D., Sánchez-Lavega, A. and Hueso, R., "Wind measurements in Saturn's atmosphere with UVES/VLT ground-based Doppler velocimetry," EPSC2016-4828 (2016).
[15] Machado, P., Valido, H., Cardesin-Moinelo, A. and Gilli, G., "Mars Atmospheric Wind Map Along the 2018 Global Dust Storm," European Planetary Science Congress, EPSC2020-221–221 (2020).
[16] Crepp, J. R., Crass, J., King, D., Bechter, A., Bechter, E., Ketterer, R., Reynolds, R., Hinz, P., Kopon, D., Cavalieri, D., Fantano, L., Koca, C., Onuma, E., Stapelfeldt, K., Thomes, J., Wall, S., Macenka, S., McGuire, J., Korniski, R., et al., "iLocater: a diffraction-limited Doppler spectrometer for the Large Binocular Telescope," presented at SPIE Astronomical Telescopes + Instrumentation, 4 August 2016, Edinburgh, United Kingdom, 990819.
[17] Haffert, S. Y., Wilby, M. J., Keller, C. U. and Snellen, I. A. G., "The Leiden EXoplanet Instrument (LEXI): a high-contrast high-dispersion spectrograph," presented at SPIE Astronomical Telescopes + Instrumentation, 9 August 2016, Edinburgh, United Kingdom, 990867.
[18] Bechter, A. J., Crepp, E. B. B. J. R., King, D. and Crass, J., "A Radial Velocity Error Budget for Single-mode Fiber Doppler Spectrographs," Ground-based and Airborne Instrumentation for Astronomy VII, 248 (2018).
[19] Bento, J., Feger, T., Ireland, M. J., Rains, A., Jovanovic, N., Coutts, D. W., Schwab, C., Arriola, A. and Gross, S., "Performance and future developments of the RHEA single-mode spectrograph," presented at SPIE Astronomical Telescopes + Instrumentation, 9 August 2016, Edinburgh, United Kingdom, 99086K.
[20] Ge, J., Angel, J. R. P. and Shelton, J. C., "Optical spectroscopy with a near-single-mode fiber-feed and adaptive optics," presented at Astronomical Telescopes & Instrumentation, 9 July 1998, Kona, HI, 253–263.
[21] Schwab, C., Leon-Saval, S. G., Betters, C. H., Bland-Hawthorn, J. and Mahadevan, S., "Single mode, extreme precision Doppler spectrographs," arXiv:1212.4867 [astro-ph] (2012).
[22] Robertson, J. G. and Bland-Hawthorn, J., "Compact high-resolution spectrographs for large and extremely large telescopes: using the diffraction limit," presented at SPIE Astronomical Telescopes + Instrumentation, 5 October 2012, Amsterdam, Netherlands, 844623.
[23] Haffert, S. Y., Por, E. H., Keller, C. U., Kenworthy, M. A., Doelman, D. S., Snik, F. and Escuti, M. J., "The Single-mode Complex Amplitude Refinement (SCAR) coronagraph: II. Lab verification, and toward the characterization of Proxima b," arXiv:1803.10693 [astro-ph] (2018).

[24] Guyon, O., "Phase-induced amplitude apodization of telescope pupils for extrasolar terrestrial planet imaging," A&A **404**(1), 379–387 (2003).

[25] Lozi, J., Guyon, O., Jovanovic, N., Goebel, S., Pathak, P., Skaf, N., Sahoo, A., Norris, B., Martinache, F., N'Diaye, M., Mazin, B., Walter, A. B., Tuthill, P., Kudo, T., Kawahara, H., Kotani, T., Ireland, M., Cvetojevic, N., Huby, E., et al., "SCExAO, an instrument with a dual purpose: perform cutting-edge science and develop new technologies," 1070359 (2018).

[26] Madec, P.-Y., Arsenault, R., Kuntschner, H., Kolb, J., Pirard, J.-F., Paufique, J., La Penna, P., Hackenberg, W., Vernet, E., Suárez Valles, M. and Hubin, N., "Adaptive Optics Facility: from an amazing present to a brilliant future...," 1070302 (2018).

[27] Marconi, A., Abreu, M., Adibekyan, V., Aliverti, M., Allende Prieto, C., Amado, P. J., Amate, M., Artigau, E., Augusto, S. R., Barros, S., Becerril, S., Benneke, B., Bergin, E., Berio, P., Bezawada, N., Boisse, I., Bonfils, X., Bouchy, F., Broeg, C., et al., "HIRES, the high-resolution spectrograph for the ELT," arXiv e-prints **2011**, arXiv:2011.12317 (2020).

[28] Markus Kasper, Nelly Cerpa Urra, Prashant Pathak, Jalo Nousianen, Byron Engler, Cédric Taïssir Heritier, Jens Kammerer, Serban Leveratto, Chang Rajani, Miska Le Louarn, Pierre-Yves Madec, Stefan Ströble, Christophe Verinaud, Adrian Glauser, Sascha Quanz, Tapio Helin, Christoph Keller, Frans Snik, Anthony Boccaletti, et al., "PCS – roadmap for exoearth imaging with the ELT," Messenger(182) (2020).

[29] Turbet, M., Leconte, J., Selsis, F., Bolmont, E., Forget, F., Ribas, I., Raymond, S. N. and Anglada-Escudé, G., "The habitability of Proxima Centauri b. II. Possible climates and observability," Astronomy and Astrophysics **596**, A112 (2016).